# Modeling nanoconfined reaction kinetics:

# Alternative methodology incorporating equilibrium extent fluctuations

Leonid Rubinovich and Micha Polak

*Department of Chemistry, Ben-Gurion University of the Negev, Beer-Sheva, Israel*

This study reveals that Equilibrium Constant Differential Equations (ECDE) for nanoconfined reactions derived recently in the frameworks of statistical mechanics are useful in modeling stochastic chemical kinetics. It is assumed and verified that if the transient value of the nanoreaction quotient is treated as being an "equilibrium constant", the corresponding "equilibrium reaction extent" and its fluctuations (the variance function) coincide with the respective transient values. The ECDE-computed variance function facilitates the solution of the stochastic kinetics equations (SKE), as is demonstrated for a stoichiometric exchange reaction. The results obtained by this original methodology are in full agreement with those provided by the chemical master equations. Contrary to the commonly used approaches based on the latter and the Gillespie algorithm, which need a lot of computer memory and are time-consuming, the proposed SKE-ECDE method requires to solve only the ECDE and a single stochastic kinetics differential equation.

Chemical-equilibrium involving a small number of molecules inside a confined nanospace can exhibit considerable deviations from the macroscopic thermodynamic limit (TL) due to reduced mixing entropy, as was predicted in several of our works using statistical-mechanics canonical partition-functions[1-3]. Our previous work[4] substantially advanced studies of the "nanoconfinement entropic effect on chemical equilibrium" (NCECE) by focusing on variations of the ordinary Equilibrium Constant Equations (ECE) due to nanoconfinement. It presented derivation of Equilibrium Constant Differential Equations (ECDE) for the reaction extent, $\xi^{eq}$, as a function of equilibrium constant, $K$, which are distinct from the conventional ECEs for the TL case. This work reveals a linkage between the ECDE and nanoconfined stochastic chemical kinetics.

The time dependence of the stochastic nanoconfined reaction extent differs from the corresponding kinetics of the macroscopic system[2,3], and as is well-known the equilibrium extent can be obtained from the kinetics in the limit of long times. The inverse possibility of modeling nanoconfined stochastic chemical kinetics using the ECDE-computed variance function is explored here in the case of the stoichiometric exchange reaction,

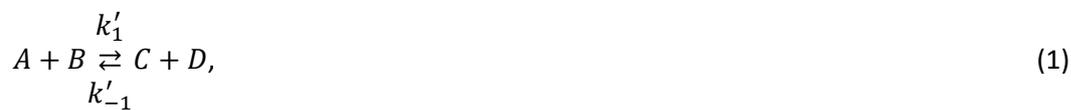

$$A + B \underset{k'_{-1}}{\overset{k'_1}{\rightleftarrows}} C + D, \tag{1}$$

where, $k'_1$ and $k'_{-1}$ denote the forward and backward elementary reaction rate constants, respectively. The number of product molecules defines the "reaction extent", $x$, as $N_C = N_D = Nx$, and the number of reactant molecules $N_A = N_B$ is $N(1-x)$, with $N$ denoting the initial (maximal) number of $A$ and $B$ molecules.

According to our previous publication[2], the stochastic kinetics equation (SKE) for the average reaction extent, $\xi \equiv \langle x \rangle$, includes the reaction extent thermal fluctuation, $\sigma^2(\xi) \equiv \langle x^2 \rangle - \xi^2$,

$$\frac{d\xi}{dt} = k'_1(1-\xi)^2 - k'_{-1}\xi^2 + (k'_1 - k'_{-1})\sigma^2(\xi). \tag{2}$$

Since the "variance function" $\sigma^2(\xi)$ is generally unknown, in Ref.[2] the equations were analyzed only qualitatively and the modeling employed the Gillespie algorithm (GA) of time-consuming dynamic Monte Carlo simulations. (To obtain a description of the kinetics corresponding to the exact solution of the Chemical Master Equations, CME, the algorithm has to be repeated many times.)

In the present study a "nanoreaction quotient" is introduced and considered as a useful concept for finding the linkage between the equilibrium and the time-dependent variances. Similarly to the commonly known macroscopic reaction quotient, the nanoreaction quotient is defined by the same formula as the equilibrium constant[4] except that in this case the molecule numbers depend on time, namely, for the stoichiometric exchange reaction,

$$Q(t) = \frac{\langle N_C(t)N_D(t) \rangle}{\langle N_A(t)N_B(t) \rangle} = \frac{(\xi(t))^2 + \sigma^2(t)}{(1-\xi(t))^2 + \sigma^2(t)}. \tag{3}$$

The mathematical identity of the expressions for the quotient and for the equilibrium constant links the reaction kinetics and equilibrium in the macroscopic case for which the variance is negligible. In particular, if the transient value of $Q(t) = \frac{\xi(t)^2}{(1-\xi(t))^2}$ is considered as "equilibrium constant", the corresponding "equilibrium reaction extent" would coincide with the transient $\xi(t)$. Furthermore, a similar conclusion is valid in the other limiting case of the smallest nanosystem containing only one pair of molecules ($N = 1$). This nanosystem is characterized by a single independent probability, $\xi = p_1$, that all molecules are products (the reactant probability is $p_0 = 1 - \xi$). The following variance function is obtained both in equilibrium and before,

$$\sigma^2(\xi) = p_0 \xi^2 + p_1(1-\xi)^2 = (1-\xi)\xi^2 + \xi(1-\xi)^2 = \xi(1-\xi). \tag{4}$$

Substituting Eq.4 in Eq.3 gives again mathematically identical dependences of the quotient and of the equilibrium constant on the reaction extent, namely,

$$Q(t) = \frac{\xi(t)}{1-\xi(t)}, \tag{5}$$

and

$$K = \frac{\xi^{eq}}{1-\xi^{eq}}. \tag{6}$$

Again, if the transient value of $Q(t)$ is considered as "equilibrium constant", the corresponding "equilibrium reaction extent" would coincide with the transient $\xi(t)$, and accordance with Eq.4 the variance at equilibrium would coincide with the transient value $\sigma^2(t)$.

Based on its validity in the two limiting cases (the macroscopic system and the smallest system), a similar argument regarding the coincidence of the respective transient and equilibrium characteristics is assumed and verified also for $N > 1$ nanosystems. Namely, every couple of values $\xi(t)$ and $\sigma^2(t)$ corresponds to a state characterized by the "equilibrium constant" $K' = Q(t)$ and by $\xi(K') = \xi(t)$ with $\sigma^2(K') = \sigma^2(t)$. Hence, the same dependence $\sigma^2(\xi)$ should correspond to the couple of kinetics functions $\xi(t)$ and $\sigma^2(t)$, as well as to the couple of equilibrium functions $\xi(K')$ and $\sigma^2(K')$. Thus, the latters can be obtained by solving the corresponding ECDE[4],

$$\frac{1}{N}(K-1)\frac{\partial \xi(K)}{\partial \ln K} = \big(\xi(K)\big)^2 - K\big(1-\xi(K)\big)^2. \tag{7}$$

and by the differentiation[4],

$$\sigma^2(K) = \frac{1}{N}\frac{\partial \xi(K)}{\partial \ln K}. \tag{8}$$

Then, the obtained $\xi(K)$ and $\sigma^2(K)$ can provide the variance function $\sigma^2(\xi)$.

The validity of the variance function $\sigma^2(\xi)$ both at equilibrium and before facilitates stochastic kinetics computations according to the following procedure shown in the flow diagram (Fig.1). Firstly, the ECDE Eq.7 is solved and $\xi(K)$ is obtained. Then, a numerical differentiation (Eq.8) provides the variance, which is substituted into the SKE (Eq.2) that is solved numerically, yielding the nanoconfined stochastic kinetics $\xi(t)$.

The computations unequivocally confirm the above assumption regarding the variance function, since the obtained kinetics $\xi(t)$ (Fig.2a) is in full agreement with a numerical solution of the exact CME. A clear system size effect on kinetics can be noted, namely, the fastest rates are obtained for the smallest system ($N = 1$), while the $N = 10$ plot is quite close to the macroscopic system behavior. The rate enhancement is associated to the positive fluctuation-related variance function $\sigma^2(\xi)$ in the SKE (Eq.2)[2]. The reaction extent enhancement for $N = 1$ exhibits a maximum (Fig.2a, inset) in the region of maximal $Q(t)$ slope (Fig.2b) close to $t \approx 0.2$. It is related to the faster equilibration in the nanosystem compared to the macrosystem, so the kinetic effect can transiently become even higher than the NCECE.[2] According to Fig.2b the nanoreaction quotient $Q(t)$ monotonously increases approaching the value of the reaction equilibrium constant $K = \frac{k_1'}{k_{-1}'} = 10$. The variance exhibits a maximum close to $t \approx 0.1$ when $\sigma^2 = 0.25$ and $\xi = 0.5$ (Fig.2a). This maximum position is fully consistent with the variance function given by Eq.4.

To conclude, the reaction extent variance making the difference between stochastic and deterministic kinetics equations can be obtained by the ECDE computations. Thus, the revealed linkage between nanoconfined equilibrium and stochastic kinetics facilitate the original methodology of modeling the kinetics, which is verified by the full agreement of the predicted results with the solutions of the chemical master equations in the case of stoichiometric exchange reactions. The computational advantages of the proposed method over the commonly used CME and GA approaches are distinct. In particular, it requires to solve only the ECDE and a single stochastic kinetics differential equation, whereas in the CME approach the number of differential equations is of the order of the number of molecules, and the GA computations have to be repeated many times in order to remove Monte Carlo simulation noise. As will be shown elsewhere, the introduced SKE-ECDE methodology is universally valid for other reactions.

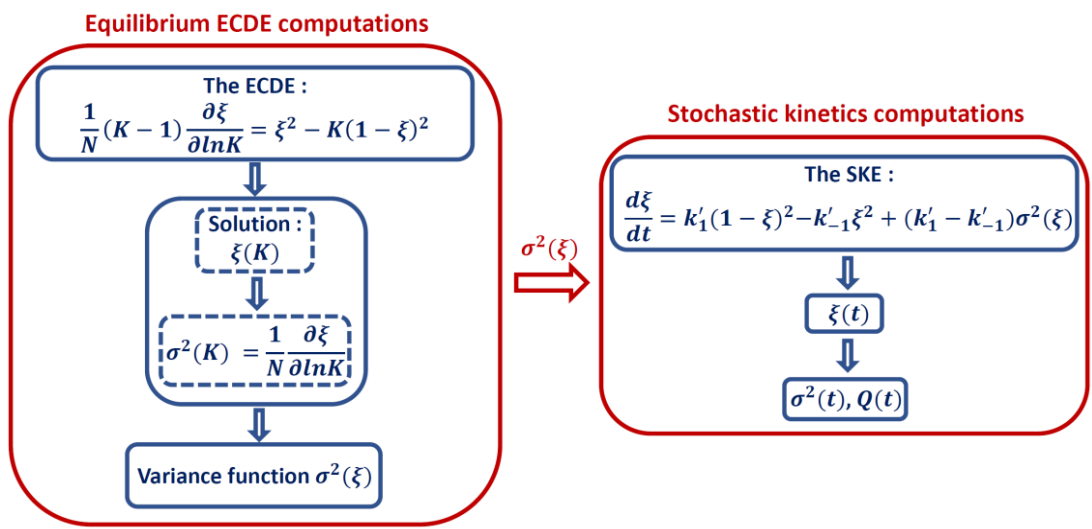

Fig. 1. Flow diagram of stochastic kinetics computations (by the MATLAB package) for the stoichiometric exchange reaction $A + B = C + D$.

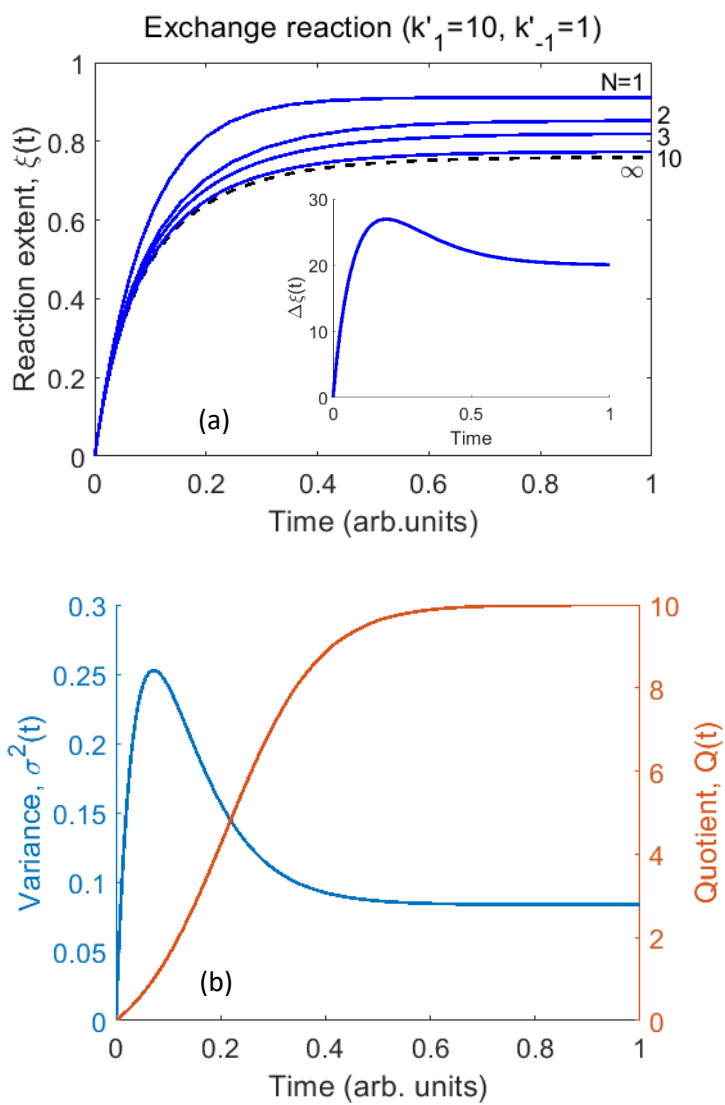

Fig. 2. (a) Stochastic kinetics of the nanoconfined exchange reaction $(N = 1 - 10)$ $A + B = C + D$ in comparison to the deterministic kinetics $(N = \infty)$. The small system computed plots coincide with those obtained by direct solutions of the Chemical Master Equations (CME). Inset: Relative enhancement of the reaction extent $\Delta \xi \equiv \left(\frac{\xi}{\xi^\infty} - 1\right) 100\%$ for $N = 1$. (b) Plots of the time-dependent variance and of the nanoreaction quotient for $N = 1$.